\documentclass[5p]{elsarticle}

\usepackage{definitions}

\journal{Nuclear Instruments and Methods in Physics Research: Section A}

\begin{document}

\begin{frontmatter}

\title{Test beam studies of the light yield, time and coordinate resolutions of scintillator strips with WLS fibers and SiPM readout
  }

\author[fnal]{Dmitri Denisov}
\author[ihep]{Valery Evdokimov}
\author[vinca]{Strahinja Lukić\corref{corrauth}}
\ead{slukic@vinca.rs}
\author[vinca]{Predrag Ujić}

\address[fnal]{Fermilab, Batavia IL, USA}
\address[ihep]{Institute for High Energy Physics, Protvino, Russia}
\address[vinca]{Vinča Institute, University of Belgrade, Serbia}
\cortext[corrauth]{Corresponding author}

\begin{abstract}
Prototype scintilator+WLS strips with SiPM readout for large muon detection systems 
were tested in the muon beam of the Fermilab Test Beam Facility. Light yield of up 
to 137 photoelectrons per muon per strip has been observed , as well as time resolution 
of 330~ps and position resolution along the strip of 5.4~cm. 
\end{abstract}


\end{frontmatter}

\section{Introduction}
\label{sec:intro}

In our previous paper \cite{Denisov16}, several designs of scintillator strips for the muon system of a detector at a future collider experiment are described, the requirements on the time and position resolution outlined, and studies of prototype strips with cosmic rays are presented. Scintillator strips for a muon system are expected to be several meters long, $\sim 1\unit{cm}$ thick and several cm wide. They include wavelength-shifting (WLS) fibers to transport the light to both ends of the strip to be read out by a pair of silicon photomultipliers (SiPM). In this paper, studies of such scintillator strips in the muon beam at the Fermilab Test Beam Facility (FTBF) are presented using state-of-the-art low-noise SiPM and measurement settings designed to reach and measure the ultimate time and coordinate resolution with various strip-readout configurations.

The primary beam at FTBF consists of 120~GeV/c protons with particle intensity up to 300~kHz. For our tests this beam is used to create a secondary beam of pions. A secondary beam with particle momenta of 28~GeV/c is selected using dipole magnets. The muon beam is produced in flight by decay of pions, generating a broad distribution of muon momenta from 16 to 28~GeV/c. 183~m downstream from the momentum-selection dipole, a 3.2~m thick concrete absorber removes all beam particles except the muons. Muon fluxes achieved with this beam are about one hundred times higher than the cosmic radiation flux. This allows accumulation of statistics within a few minutes per data point. The $1\,\sigma$ radius of the muon beam after the concrete wall is measured to be 5~cm, with a $1\,\sigma$ angular spread of $\sim 3\degrees$ \cite{muonBeam}.

Intrinsic limits of the timing precision of SiPMs have been explored elsewhere using scintillators of several mm in size with direct SiPM readout \cite{lyso16, Stoykov2012}. Although different in size than the devices tested in our studies, these results demonstrate the excellent potential of the SiPM for time measurements.
In this work, Hamamatsu S13360-3050CS SiPMs are used for light detection \cite{S13360-3050CS}. The low noise level of these SiPMs allows setting discriminator thresholds below the one-photoelectron signal amplitude level, thus effectively measuring the arrival time of the first photon. 

The test setup used is described in \secref{sec:setup}, the tested strips are described in \secref{sec:strips}, results are discussed in \secref{sec:results} and conclusions are given in \secref{sec:conclusions}.

\section{Setup}
\label{sec:setup}

\begin{figure}
\centering
   \includegraphics[width=\figwid]{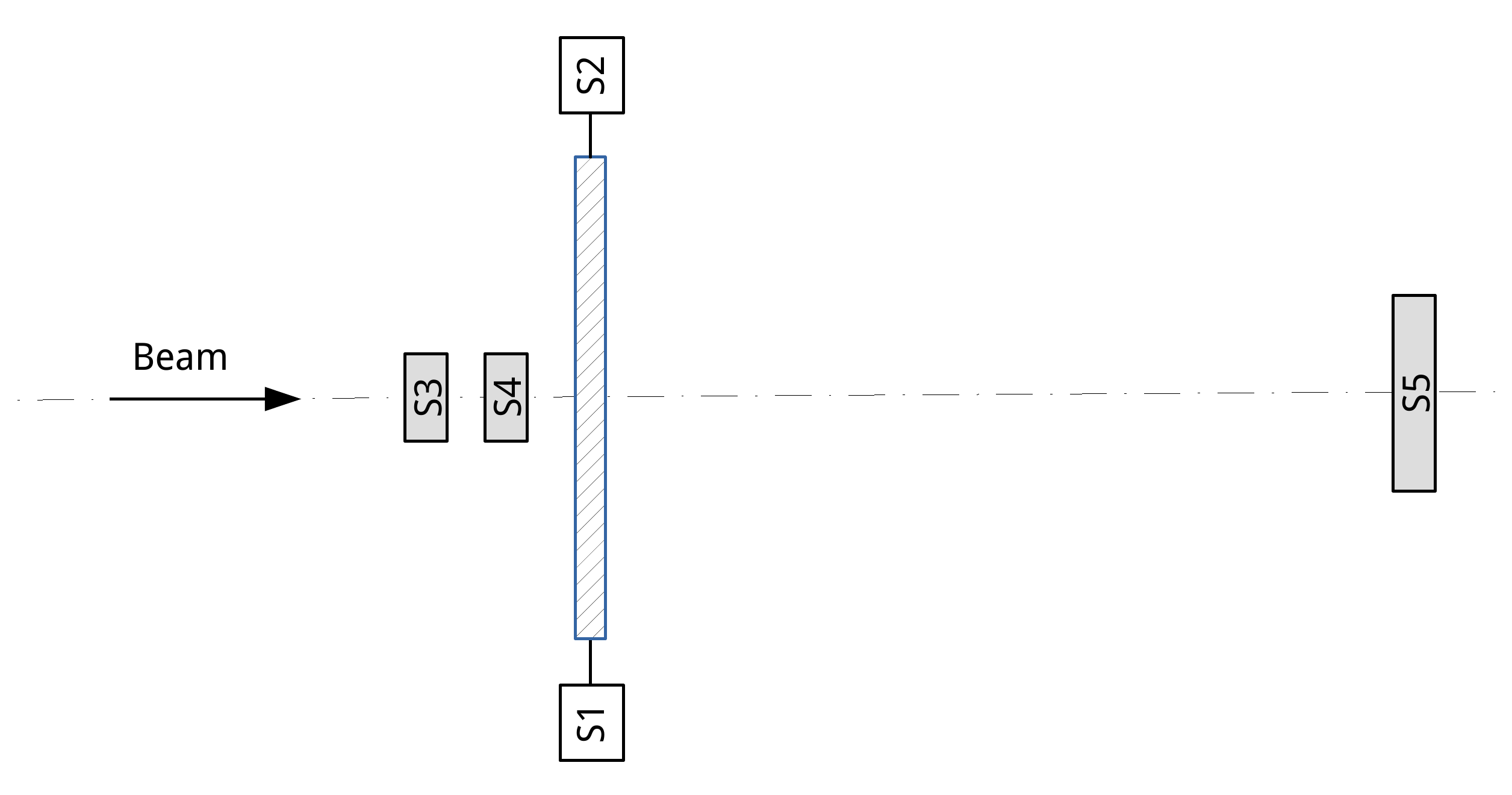}
   \caption{\label{fig:setup} Measurement setup seen from above (not to scale). The dash-dotted line represents the beamline.}
\end{figure}

The measurement setup is shown in Fig.\ \ref{fig:setup}.
\sthr and \sfour are \bicron scintillator strip counters with a $27\times12\unit{mm}^2$ profile and vacuum photomultiplier tube (PMT) readout. The length of \sthr is 15~cm and the length of \sfour is 40~cm. Both \sthr and \sfour are installed vertically, with the bottom end 5~cm below the beam center and the PMT connected to the bottom end. \sfive is a scintillation counter with an area of $16\times24\unit{cm}^2$, thickness of 12~mm and a vacuum PMT readout. \sthr, \sfour and \sfive are aligned along the beam line. The distance between \sthr and \sfour is 10~cm and the distance between \sfour and \sfive is 95~cm. 

The tested strip is mounted on a 4~m long aluminum bar, together with the readout boxes \sone and \stwo containing the SiPMs and the preamplifiers. The aluminum bar is installed on an aluminum rail so that the bar can slide along the rail in the horizontal direction perpendicular to the beam. In this way different positions of muon impact, $x$, are scanned along the tested strips. The strip and the boxes are fixed to the bar so that the fibers connecting them do not move during the scans. 
To measure $x$ during the scans an adhesive tape with distance marks at 1~cm pitch is affixed to the rail, and a reference position is marked on the sliding bar on which the tested strip is mounted. The reading $x=0$ corresponds to the position of the bar in which the longitudinal center of the tested strip is located against the center of the trigger counters \mbox{\sthr} and \mbox{\sfour}.

A CAMAC system with a LeCroy 2249A 12-input charge-sensitive ADC \cite{LeCroyADC} and a LeCroy 2228A 8-input TDC \cite{LeCroyTDC} is used to digitize the amplitude and the arrival time of the signals. 

The data collection is triggered by a coincidence between \sthr, \sfour and \sfive. 
The signal amplitudes of the counters \sthr and \sfour are recorded for the offline selection of muons by restricting the analysis to events with at least 80\% of the most probable muon energy deposit in both \sthr and \sfour. 

The signals from \sipmo and \sipmt are each split into two circuits using passive splitters. The signal in the time circuit is amplified using a Phillips Scientific model 776 fast $10\times$ amplifier \cite{Phillips776} before the input of the constant-threshold discriminator. 
The signal in the amplitude circuit is attenuated using passive attenuators to match the dynamic range of the ADC.

\begin{figure}
\centering
   \includegraphics[width=\figwid]{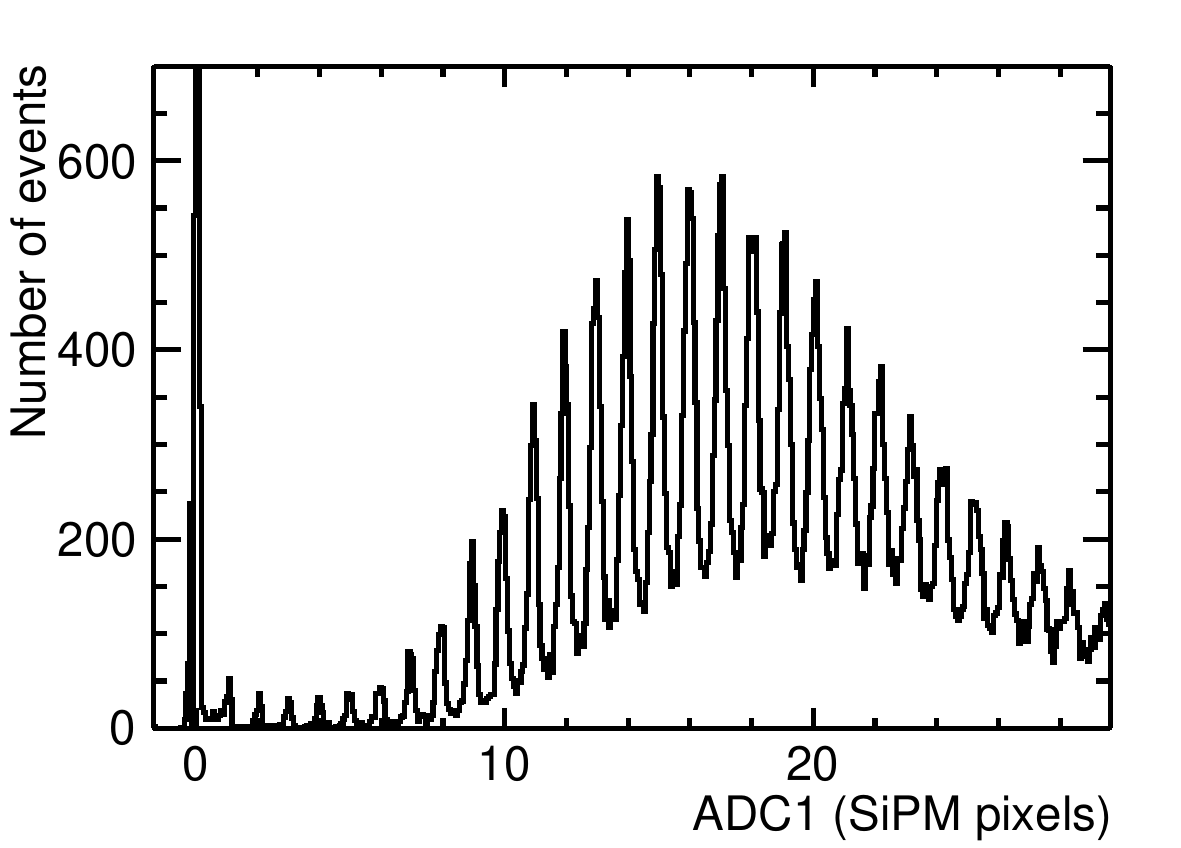}
   \caption{\label{fig:spectrum} ADC spectrum from a SiPM in the low gain mode mounted on one of the tested strip configurations from an overnight run with cosmic muons. Excellent pixel uniformity and low noise result in clear peak structure corresponding to integer numbers of firing pixels.}
\end{figure}

For each tested strip configuration, a longitudinal scan of the irradiation position is performed in two gain modes:
In the \emph{low gain} mode, the bias voltages for the SiPMs are set to $\sim 3\unit{V}$ above the breakdown voltage. This mode is characterized by a high uniformity of gain between pixels in the SiPM, allowing calibration of the single-pixel amplitudes from the amplitude spectrum of the SiPM (\figref{fig:spectrum}). The discriminator threshold for the SiPM time signals is set to $U_{\text{thr}} = 0.6\,U_{\text{pix}}$, where $U_{\text{pix}}$ is the amplitude of the signal produced by a single SiPM pixel. In this way the recorded time corresponds to the arrival of the first photon.

In the \emph{high gain} mode, the bias voltages are set at $\sim 7\unit{V}$ above the SiPM breakdown voltage. In the high gain mode the photon detection efficiency is 40\% higher than in the low gain mode and is close to saturation for this type of SiPM \cite{S13360-3050CS}. The increased photoelectron yield provides better time resolution. The crosstalk between pixels is also higher by several percent in the high gain mode. Because of the higher dark count rate of the SiPMs in the high gain mode, the discriminator threshold is set to $\sim 1.3\,U_{\text{pix}}$, effectively corresponding to the arrival of the second photon.

\section{Scintillator strip designs}
\label{sec:strips}

\begin{figure*}
  \centering
  \begin{subfigure}{0.7\figwid}
    \includegraphics[width=\textwidth]{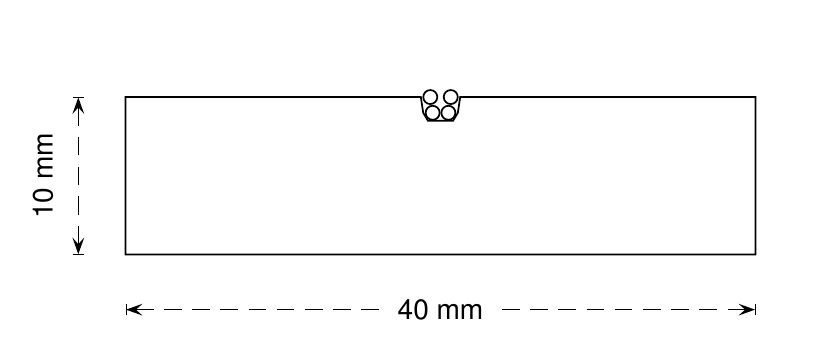}
    \subcaption{\label{fig:minos}}
  \end{subfigure}
  \begin{subfigure}{0.7\figwid}
    \includegraphics[width=\textwidth]{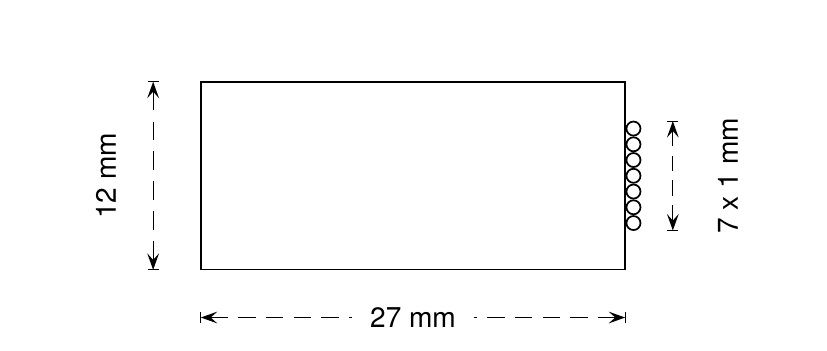}
    \subcaption{\label{fig:bicron}}
  \end{subfigure}
  \caption{\label{fig:designs} Tested designs of 
              the scintillator strips with WLS fibers: 
              \subref{fig:minos} MINOS strip with four Bicron WLS fibers,
              \subref{fig:bicron} Bicron strip with seven Bicron WLS fibers.
          }
\end{figure*}

Two designs of the scintillator strip with WLS fibers are tested, corresponding to the best-performing designs from Ref.\ \cite{Denisov16}, and schematically presented in Fig.\ \ref{fig:designs}. The description of the studied strips, fibers and light insulation is the following.

Design \subref{fig:minos} uses clear polystyrene scintillator strips with a $40\times10\unit{mm}^2$ cross section with a central groove, co-extruded with a $\text{TiO}_2$ loaded surface layer such as used for the MINOS detector \cite{MINOS_sci_08}. The inner surface of the groove is not covered with the reflective layer. Four \bicronwls WLS fibers of 1.0~mm diameter \cite{BicronWLS} are inserted into the groove and covered with white \tyvek sheet type 1056D \cite{dupont, Abazov05}. The capacity of the groove to accomodate WLS fibers is thus fully exploited. The strip is then wrapped in several layers of black \tedlar paper \cite{dupont, Abazov05}. 

Design \subref{fig:bicron} uses clear \bicron fast scintillator strip with a $27\times12\unit{mm}^2$ cross section \cite{BicronStrip}. Seven \bicronwls WLS fibers of 1.0~mm diameter are attached to the narrow side of the strip using adhesive tape. In this design, the number of WLS fibers is limited by the sensitive surface of the used photodetectors. The strip is then wrapped with one layer of the \tyvek sheet and several layers of black \tedlar paper.

In both designs, the ends of the fibers extend 20~cm beyond the end of the strip and are bundled together. In the design \subref{fig:minos}, the four fibers are bundled in a square shape, and in the design \subref{fig:bicron} the seven fibers are bundled so that six fibers surround one in a tight hexagonal shape. Mechanical connectors are used to precisely position the end of the fiber bundle in front of the SiPM for efficient light collection. No optical glues or greases are used between the fibers and the scintillator strips or the SiPMs. The SiPMs have an active surface area of $3\times3\,\unit{mm}^2$ and fully cover the fiber bundles.

The tested strips include two strips of design \subref{fig:minos}, one with the length of 1~m (strip $A$) and one with the length of 2~m (strip $B$), and one strip of design \subref{fig:bicron}, with the length of 1~m (strip $C$). 

\section{Results}
\label{sec:results}

The analysis of all measurements is performed using the following off-line selection criteria:

\begin{itemize}
  \item Presence of a signal is required in both SiPMs (TDC receives a stop signal before counting to end-of-scale)
  \item Energy deposit in each of the counters \sthr and \sfour is at least 80\% of the most probable deposit for a muon.
\end{itemize}

\subsection{Photoelectron yield per muon}
\label{sec:phyield}

To measure the photoelectron yield per muon, first the ADC spectrum is calibrated in the number of SiPM pixels using the Fourrier transform of the spectrum in the low gain mode (\figref{fig:spectrum}). Then the average of the spectrum is calculated using the selection cuts outlined above. Finally the result is corrected for the cross-talk factor. The cross-talk factor for the Hamamatsu S13360-3050CS SiPMs is only 4\% in the low gain setting, and has low sensitivity to the bias voltage or temperature. Thus it is taken from the SiPM datasheets \cite{S13360-3050CS} and not separately measured. 

In the high gain setting, the integer pixel peaks are not visible in the amplitude spectrum. This can be explained by worse pixel gain uniformity at high gain. Consequently the photoelectron yields for the high gain setting were estimated from the yields measured in the low gain setting, using the ratio of quantum efficiencies of the SiPM in the two gain settings obtained from the datasheet \cite{S13360-3050CS}. The ratio of the quantum efficiencies from the datasheet is consistent with the measured ratio of the average signal amplitudes for the two gain settings taking into account the ratio of the single-photoelectron signal amplitudes observed on the oscilloscope, the ratio of the attenuation factors introduced by the attenuators and the ratio of the cross-talk factors for the two gain settings. In the high gain setting, the cross-talk factor for the Hamamatsu S13360-3050CS SiPMs is 10\%.

\begin{figure}
  \centering
    \includegraphics[width=\figwid]{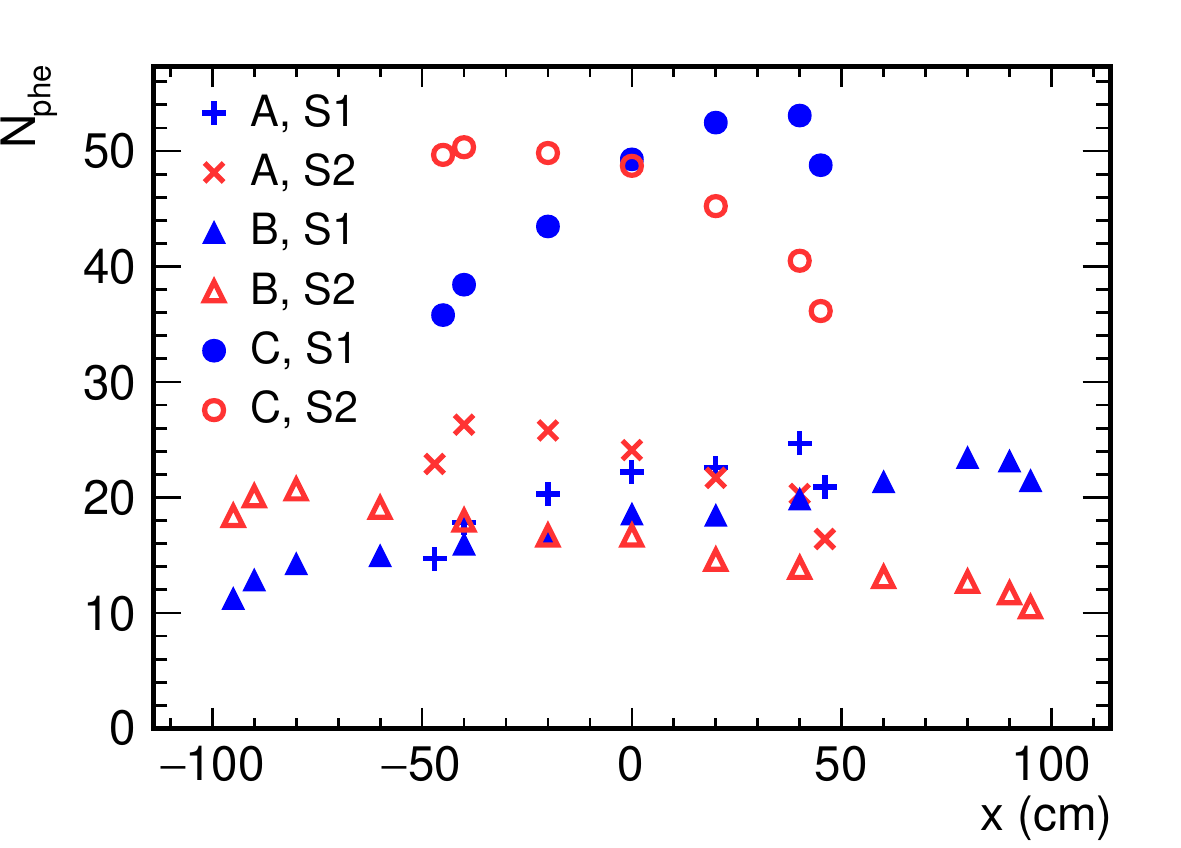}
    \caption{\label{fig:attenuation} Average number of photoelectrons per muon at each end of the strip as a function of position for all tested strip configurations ($A$, $B$, $C$) in the low gain setting. Error bars are smaller than the points on the plot.}
\end{figure}

\begin{figure}
  \centering
    \includegraphics[width=\figwid]{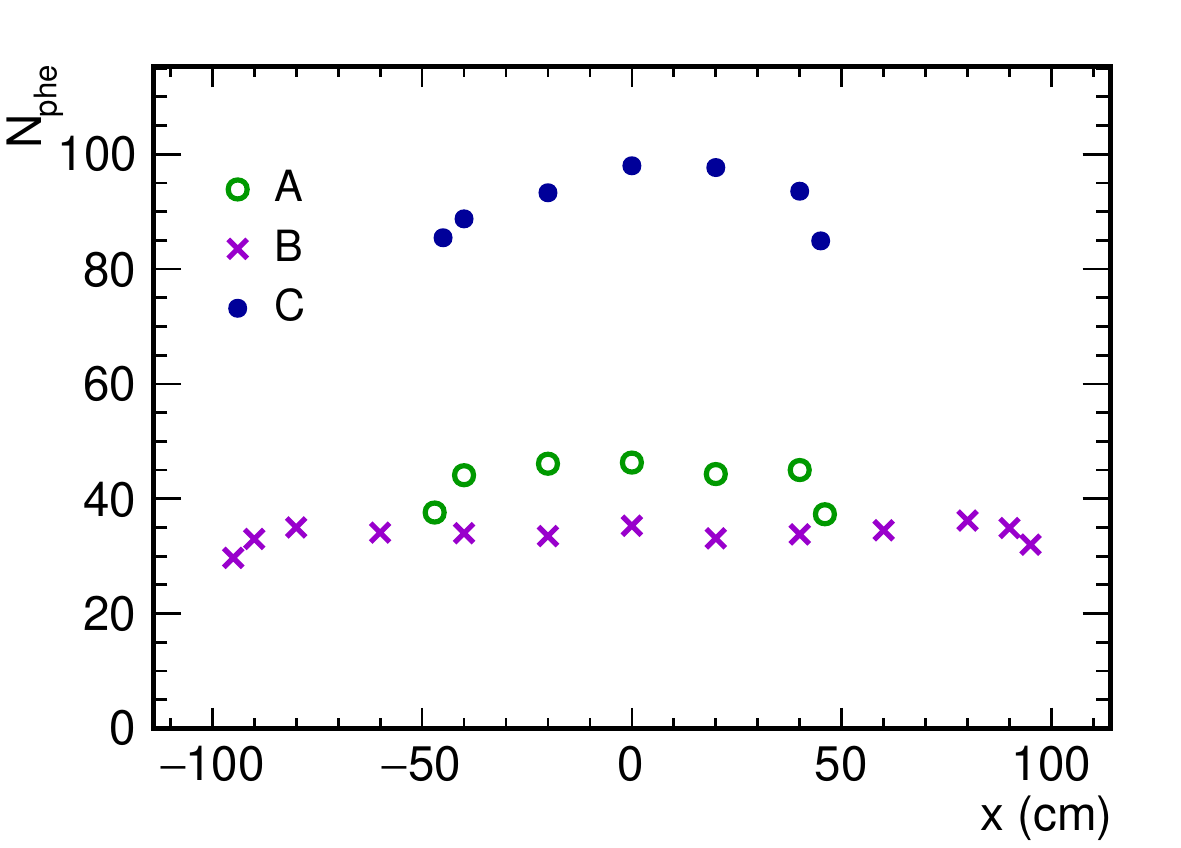}
    \caption{\label{fig:npesum} Average total number of photoelectrons per muon per strip as a function of position for all tested strips in the low gain setting. Error bars are smaller than the points on the plot.}
\end{figure}

The average photoelectron yield per muon for each end of the strip is shown in \figref{fig:attenuation} for all tested strips, as a function of the position along the strip $x$ in the low gain SiPM setting. In \figref{fig:npesum} the sum of photoelectrons from both ends is shown for the same conditions. The scatter of points for the same strip is mainly due to temperature-induced variations of the SiPM quantum efficiency of $\sim 2\%$.
The statistical uncertainties are smaller than the points on the plot.
At the distance of one strip width from each end of the strip light collection efficiency is reduced by $\sim 10\%$ because the end sides of the scintillator are not covered with reflective coating or wrapping. 
Opposite ends of the strips have slightly different performance due to the differences between individual SiPMs and minor variations in the assembly of fibers and positioning of the fiber bundles w.r.t.\ the SiPMs.

Exponential fit to the data shown in \figref{fig:attenuation} (excluding end points) allows to estimate the attenuation length. When the 2\% relative variations of photoelectron yield due to temperature are included as uncertainties of the measured points, the $\chi^2/ndf$ of the fit ranges from 1.0 to 1.7 for the four fitted curves of the strips $A$ and $B$. The $\chi^2/ndf$ of the fit for strip $C$ is 2.3 and 3.1 for the two ends of the strip, reflecting the fact that the dependence of light intensity on distance in a design consisting of a scintillator strip and WLS fibers is in general more complex than the simple exponential attenuation. Nevertheless, the exponential attenuation remains the main reason for light loss in longer strips. 
The obtained values for the attenuation length are somewhat lower than in the WLS fiber datasheet \mbox{\cite{BicronWLS}}.
Average number of photoelectrons per strip is calculated by averaging points on \figref{fig:npesum}, excluding points closest to the strip ends.
Attenuation lengths averaged from both sides of the strip and average total number of photoelectrons per muon per strip are listed in \tableref{tab:attenuation} for the three tested strips and both gain settings. 
The uncertainties for the attenuation length are obtained from the fit. The uncertainties in \ntot{} for the low gain setting include the uncertainty of the ADC calibration, the statistical uncertainty of the average number of pixels and the uncertainty of the cross-talk correction, assumed to be equal to the full size of the cross-talk factor taken from the datasheet. The uncertainties in \ntot{} for the high gain setting include in addition the uncertainty of the gain ratio between the two settings of 10\%. 

\begin{table}
\caption{\label{tab:attenuation}Average attenuation length and average total number of photoelectrons per muon \ntot{} for the three tested strips and two gain settings.}
\centering
  \begin{tabular}{l c c c}
    Strip  & $\left< \lambda \right>$ (m) & $\ntot{,LG}$ & $\ntot{,HG}$  \\
    \hline
    $A$    &  $2.8\pm0.3$  &  $46\pm2$  &  $64\pm7$ \\ 
    $B$    &  $3.3\pm0.2$  &  $35\pm1$  &  $49\pm5$  \\ 
    $C$    &  $3.1\pm0.5$  &  $98\pm5$  &  $137\pm14$ \\ 
  \end{tabular}
\end{table}

\subsection{Time resolution}

For the studies of the time resolution, the average time of the trigger counters \sthr and \sfour, $\tref = (t_3 + t_4)/2$, is used as the reference for the time of the muon passage. The resolution of \tref can be estimated from the distribution width of the variable $\Delta t_{3,4}/2 = (t_3 - t_4)/2$.
Although the discriminator threshold is set at 25\% of the most probable muon signal amplitude in \sthr and \sfour, the amplitude effect on the time measurement is not negligible. The correction of the amplitude effect for \sthr and \sfour is performed with the standard amplitude correction function $\delta t_i = C_i/A_i\; (i=3,4)$. The coefficients $C_3$ and $C_4$ corresponding to the counters \mbox{\sthr} and \mbox{\sfour} are obtained by
fitting $\Delta t_{3,4}/2$ with the two-dimensional function of $A_3$ and $A_4$

\begin{equation}
\label{eq:amp34}
  \Delta t_{3,4}/2 = \Delta t_{\infty} - C_3/A_3 + C_4/A_4.
\end{equation}

Where $A_3$ and $A_4$ are signal amplitudes of the counters \sthr and \sfour and $\Delta t_{\infty}$, $C_3$ and $C_4$ are constants.
The fitted values of $C_3$ and $C_4$ are then used to correct the amplitude effect in \sthr and \sfour, respectively. The resulting width of the $\Delta t_{3,4}/2$ distribution is between 100 and 110~ps as compared to $150\div 160\unit{ps}$ before the correction.

The distributions of the following observables are analyzed:

\begin{equation}
  \begin{aligned}
    t_1 &= TDC_1 - \tref \\
    t_2 &= TDC_2 - \tref \\
    t_{+} &= (t_1 + t_2) / 2 \\
    t_{-} &= (t_1 - t_2) / 2 
  \end{aligned}
\end{equation}
 
Here $TDC_{1,2}$ are the measured signal times from \sipmo and \sipmt. 
$t_+$ is the time of the muon passage relative to \tref as measured by the tested strip. $t_-$ is the observable used for the determination of the muon position along the strip.

\begin{figure}
\centering
   \includegraphics[width=\figwid]{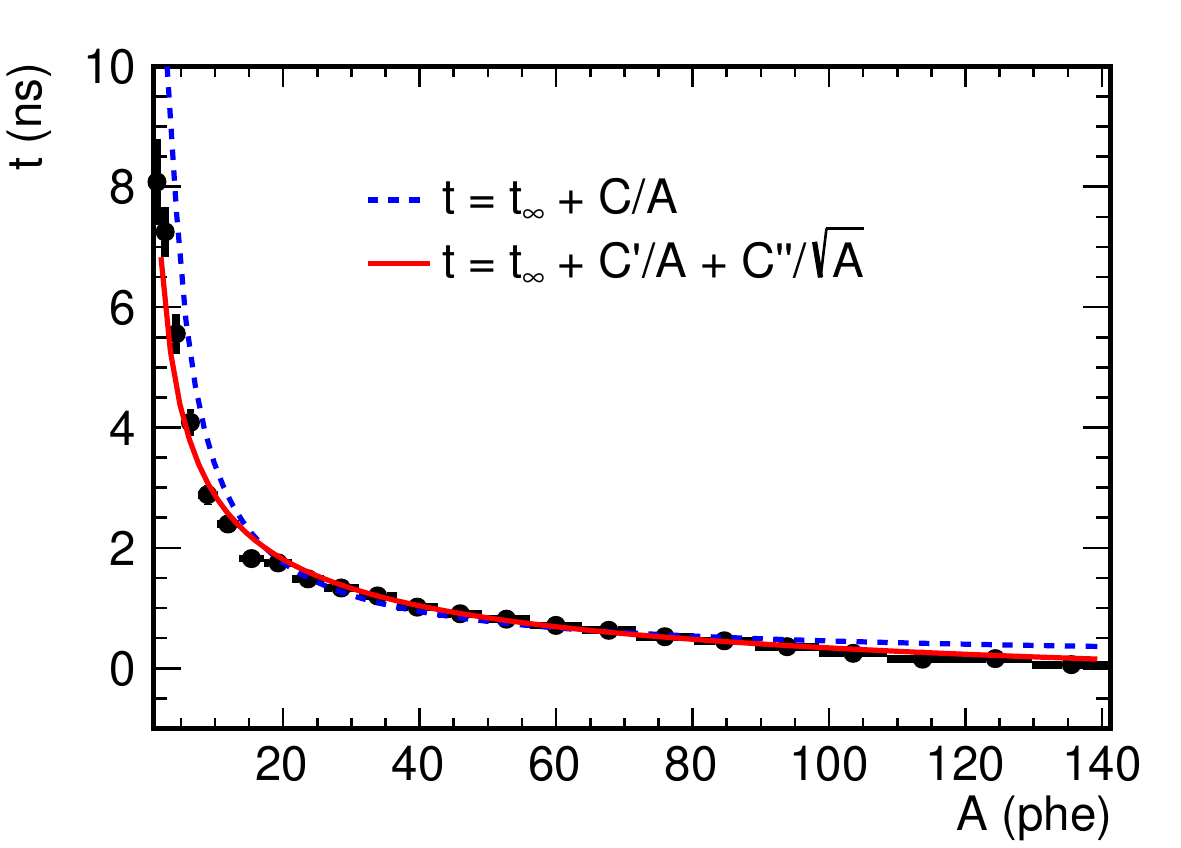}
   \caption{\label{fig:acorr12comp} Time-amplitude scatter plot for one end of the strip $C$. The fit by two functions for the amplitude correction is also shown. $A$ is the signal amplitude, $t_{\infty}$, $C$, $C'$ and $C''$ are constants.
   }
\end{figure}

The standard $1/A$ correction of the amplitude effect, derived assuming linear
dependence of the slope of the rising edge on the signal amplitude, does not describe the measured time-amplitude relationship for the tested strips well. 
The main reason for this is that the TDC of the tested strips is triggered by the arrival of a small number of photoelectrons, of the order of unity. 
In such a situation, the measured time is influenced by the statistics of the produced photoelectrons and their time distribution.
To better describe the time-amplitude relationship, an \textit{ad-hoc} term proportional to $1/\sqrt{A}$ is added to the correction function.
\mbox{\figref{fig:acorr12comp}} shows the time-amplitude scatter plot for one end of the strip $C$, together with the fit of both discussed amplitude correction functions.
The function with the additional $1/\sqrt{A}$ term fits the data much better.
The residual deviations are negligible considering the achievable time resolution with the tested strips.

To verify that the above relationship originates from the distribution of the arrival times of the triggering photoelectrons, a Monte Carlo simulation of the time distribution of the photoelectrons was performed, taking into account the time distributions of the scintillation, the fluorescence and the propagation through the WLS fiber. The simulation confirms that the above function is a good approximation of the time-amplitude relationship for the first photoelectron over a wide range of amplitudes.
Thus the amplitude correction is performed using a function of the form $t_{\infty} + C'/A + C''/\sqrt{A}$, where $A$ is the signal amplitude and $t_{\infty}$, $C'$ and $C''$ are constants.

\begin{figure}
\centering
   \includegraphics[width=\figwid]{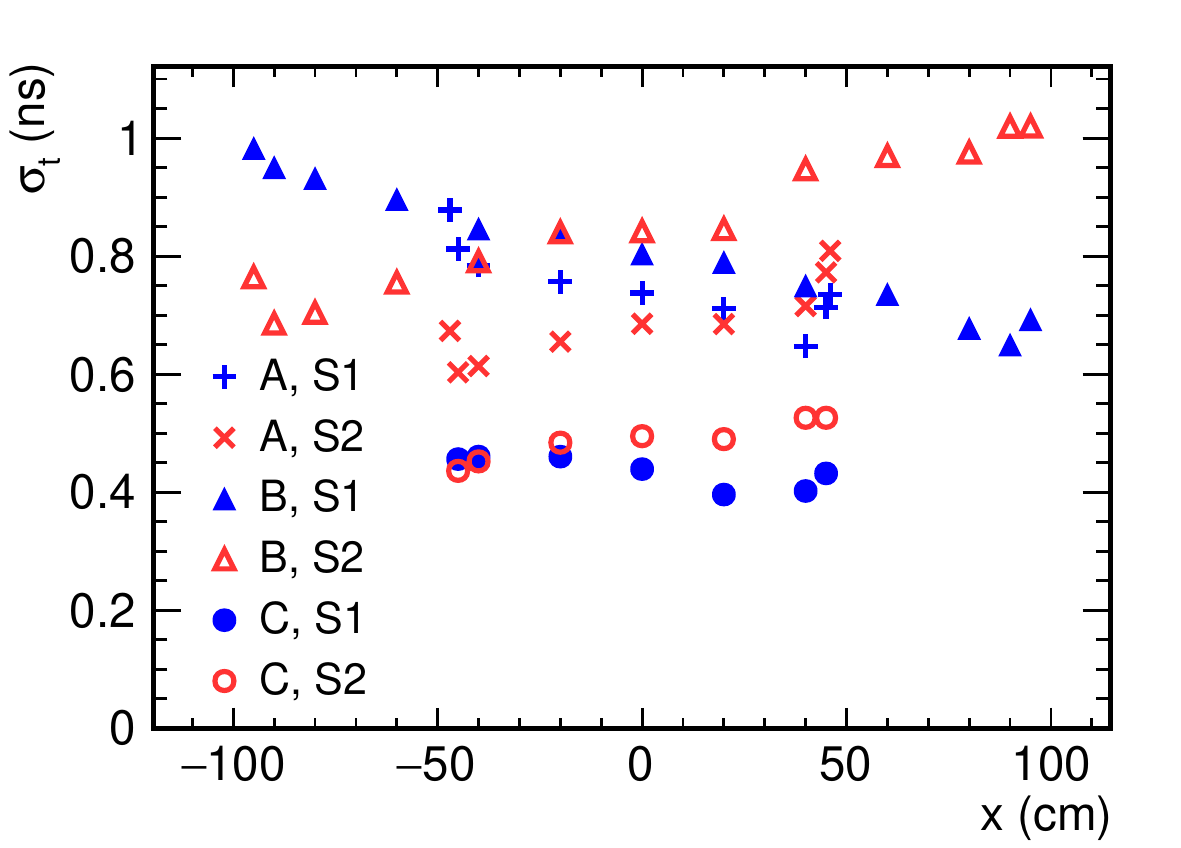}
   \caption{\label{fig:sig1sig2} Resolution of $t_1$ and $t_2$ at different positions along the tested strips. Error bars are smaller than the points on the plot.}
\end{figure}

The resolutions of $t_1$ and $t_2$ at different muon positions along the strip are shown in \figref{fig:sig1sig2} for the three tested strips in the high gain setting. All resolutions shown represent the fitted Gaussian $\sigma$ of the corresponding distribution, corrected for the resolution of \tref. As expected, the time resolution worsens with the increase in distance of the muon from the respective SiPM. For muons at a distance from the strip end smaller than one strip width additional deterioration of resolution is observed due to the poor light collection.

\begin{figure}
\centering
   \includegraphics[width=\figwid]{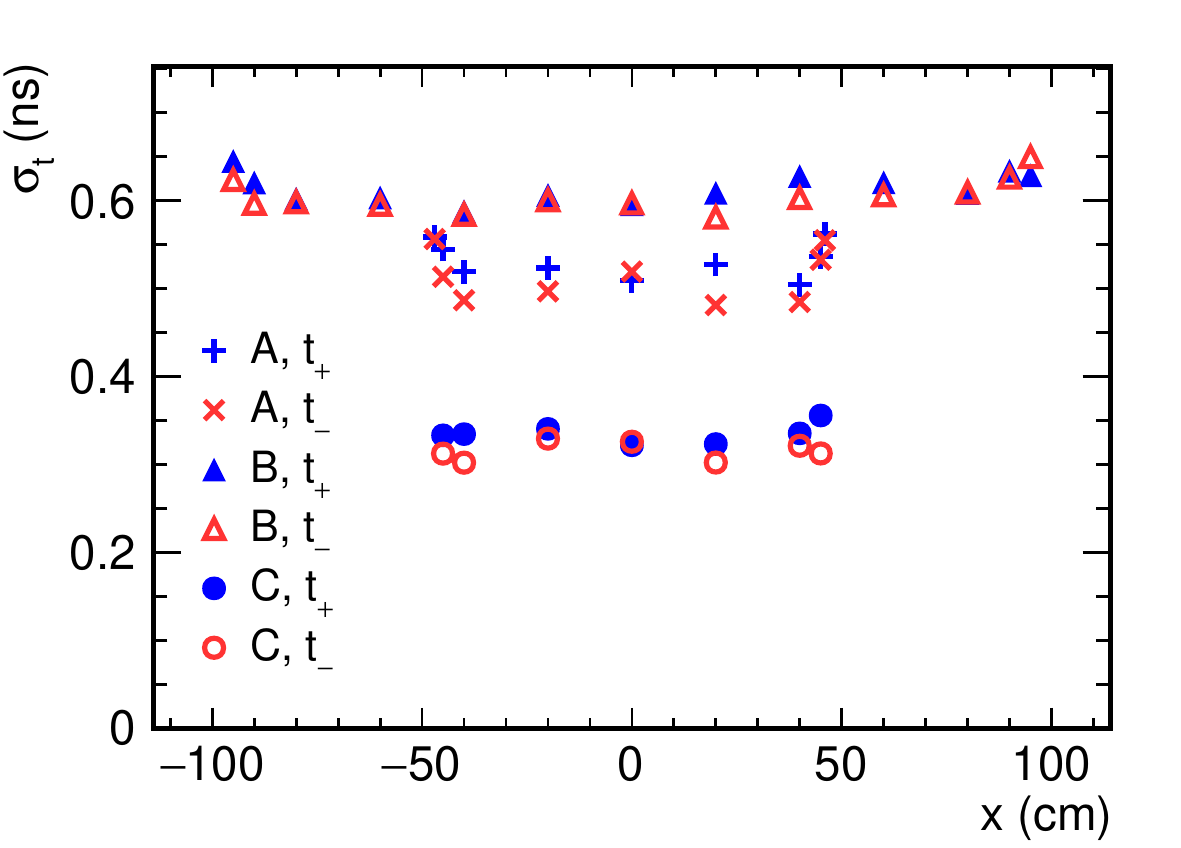}
   \caption{\label{fig:sigmt-sigdt} Resolution of $t_+$ and $t_-$ at different positions along the tested strips. Error bars are smaller than the points on the plot.}
\end{figure}

The resolutions of $t_+$ and $t_-$ at different muon positions along the three tested strips are shown in \figref{fig:sigmt-sigdt} for the high gain setting. The measured resolutions of $t_+$ and $t_-$ both contain small additional contributions from systematic effects. In the case of $t_+$, the additional spread comes from the resolution of \tref, which is about 100~ps. In the case of $t_-$, the additional spread comes from the spread of the muon positions over the zone defined by the trigger scintillators. Both these effects have been corrected by quadratic substraction. The largest relative corrections are made for the strip $C$, 5\% for the $t_+$ resolution and 1.4\% for $t_-$.

\subsection{Position resolution}

The muon position along the strip, $x$, is determined using the relationship

\begin{equation}
\label{eq:position}
  x = v^* \frac{t_2 - t_1}{2} = v^* t_-
\end{equation}

\begin{figure}
\centering
   \includegraphics[width=\figwid]{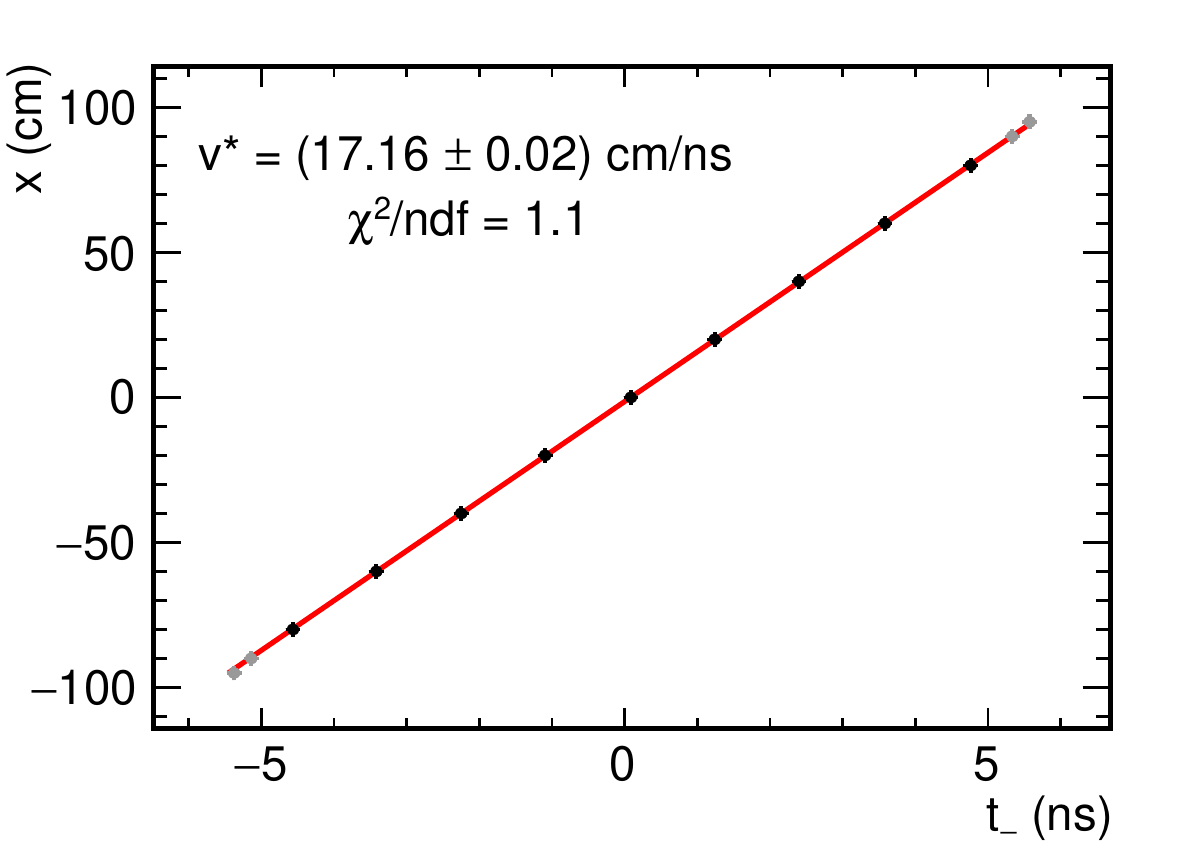}
   \caption{\label{fig:speed} Straight line fit to the data points of the muon impact position $x$ vs.\ $t_- = (t_2 - t_1)/2$ for the strip $B$ in the high gain setting. End points are excluded from the fit. Error bars of $x$ and $t_-$ are smaller than the points.}
\end{figure}

The effective speed of the signal propagation along the strip, $v^*$, is measured by 
fitting a straight line to the data points of $x$ vs.\ $t_- = (t_2 - t_1)/2$.
An example is shown in \figref{fig:speed} for the strip $B$ in the high gain setting. The end points are excluded from the fit. The uncertainties on $x$ come from the visual alignment procedure of the slider carrying the tested strip and are about 1~mm, while the uncertainties on $t_-$ are statistical uncertainties of the mean of the distribution of $t_-$ in each measurement.

\begin{figure}
\centering
   \includegraphics[width=\figwid]{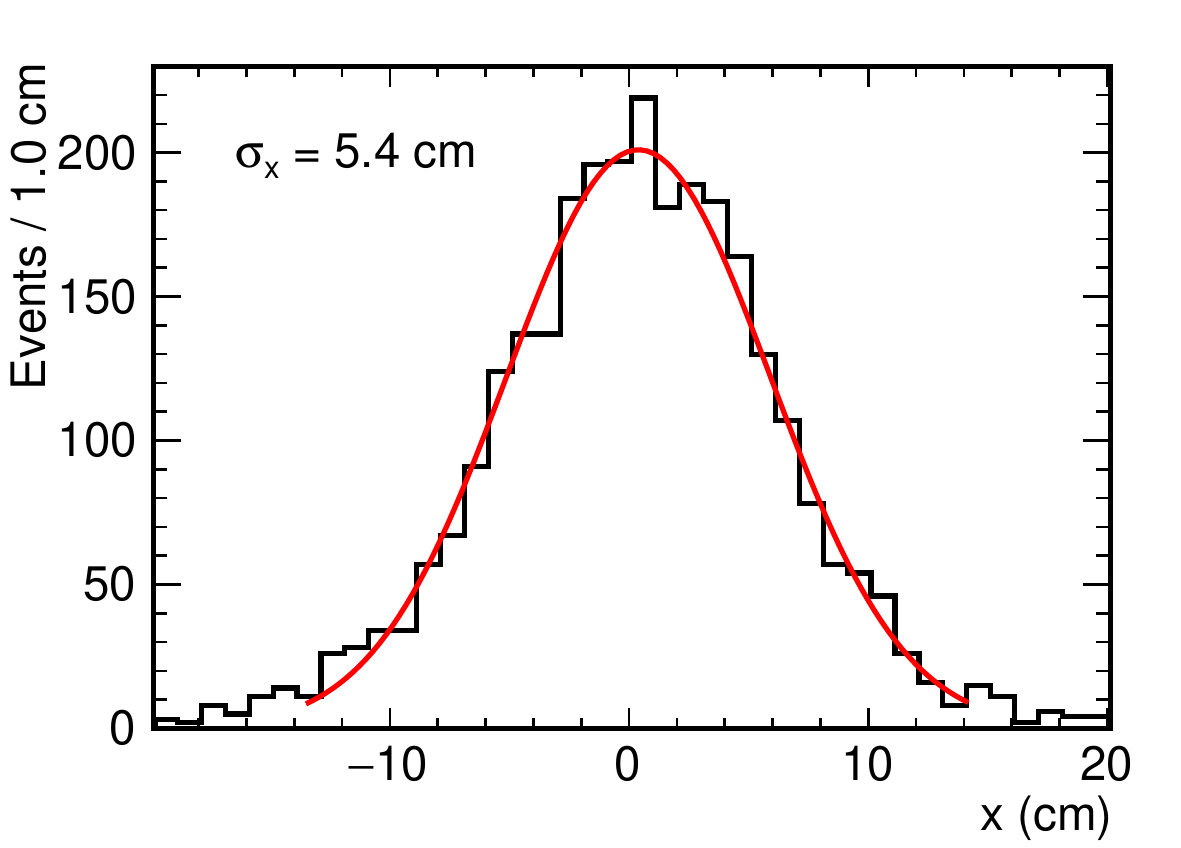}
   \caption{\label{fig:sigx} Distribution of muon hit positions determined from the time difference for the strip $C$ in the high gain setting. Beam impact is at $x=0$. Gaussian fit to the data is also shown.}
\end{figure}

\figref{fig:sigx} shows the distribution of muon positions determined using Eq.\ \eqref{eq:position} for the muon impact at $x=0$ for the case of strip $C$ with the high gain setting.
The influence of the width of the trigger counters \sthr and \sfour on the measured $\sigma_x$ was corrected by quadratic subtraction of the RMS width of the position distribution of the selected muons. The distribution of muon positions selected by the trigger counters is nearly uniform with the full width of 2.7~cm. The RMS width is thus 0.8~cm. The correction is smaller than 1~mm in all cases.

\begin{figure}
\centering
   \includegraphics[width=\figwid]{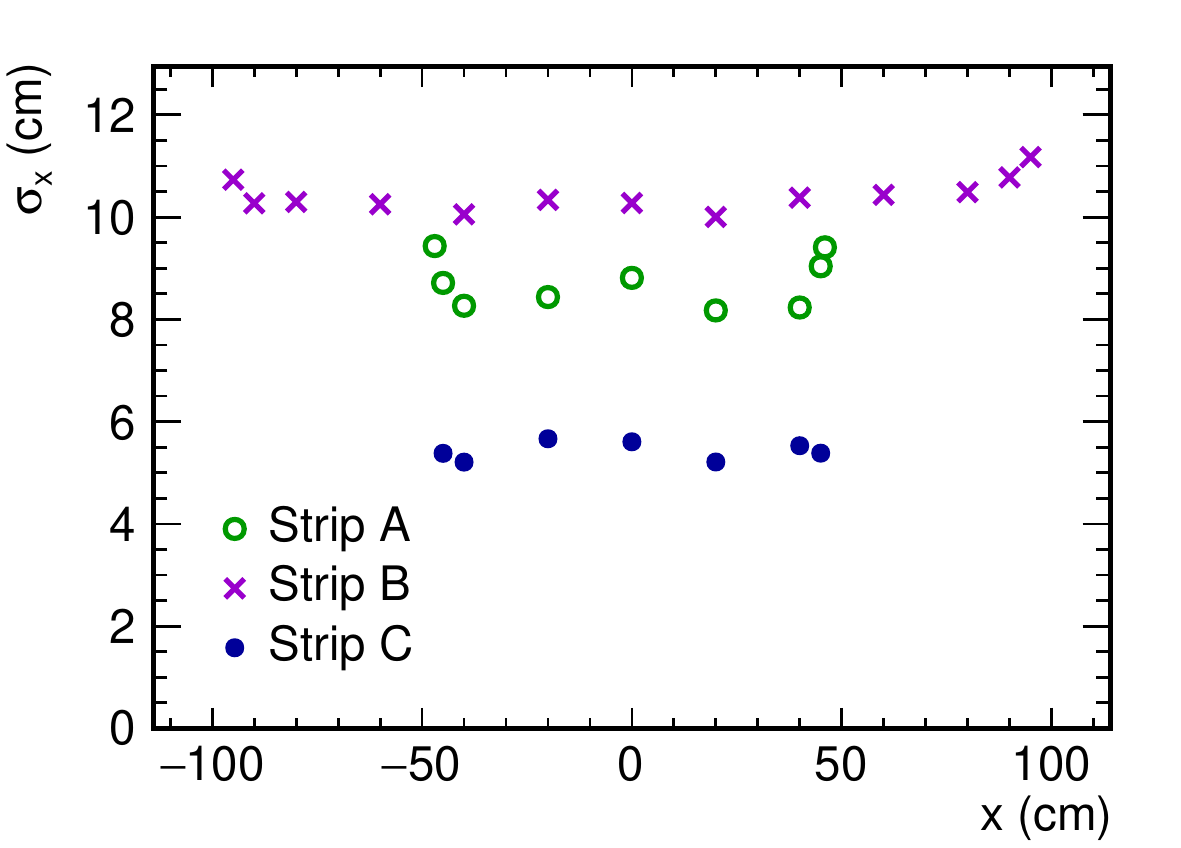}
   \caption{\label{fig:sigxall} Position resolution at different positions along the tested strips. Error bars are smaller than the points on the plot.}
\end{figure}

Position resolutions for the three tested strips in the high gain setting
are shown in \figref{fig:sigxall} for various beam impact positions. All resolutions shown represent the fitted $\sigma$ widths of the corresponding Gaussian distributions.

\subsection{Discussion of the results}

\begin{table}
\centering
\caption{\label{tab:performance} Results for all tested strips in the high gain setting.}
  \begin{tabular}{l c c c c c c}
      Strip & $\sigma_1$ & $\sigma_2$ & $\sigma_-$ & $\sigma_+$ &  $v^*$   & $\sigma_x$ \\
            &     ns     &     ns     &     ns     &     ns     &  cm/ns   &    cm     \\
    \hline
      $A$   &     0.73   &     0.67   &    0.49    &   0.52     & $16.97\pm0.04$ &    8.3     \\
      $B$   &     0.81   &     0.85   &    0.60    &   0.61     & $17.16\pm0.02$ &   10.3     \\
      $C$   &     0.45   &     0.46   &    0.32    &   0.33     & $16.91\pm0.05$ &    5.4    \\
  \end{tabular}
\end{table}

\begin{figure}
\centering
   \includegraphics[width=\figwid]{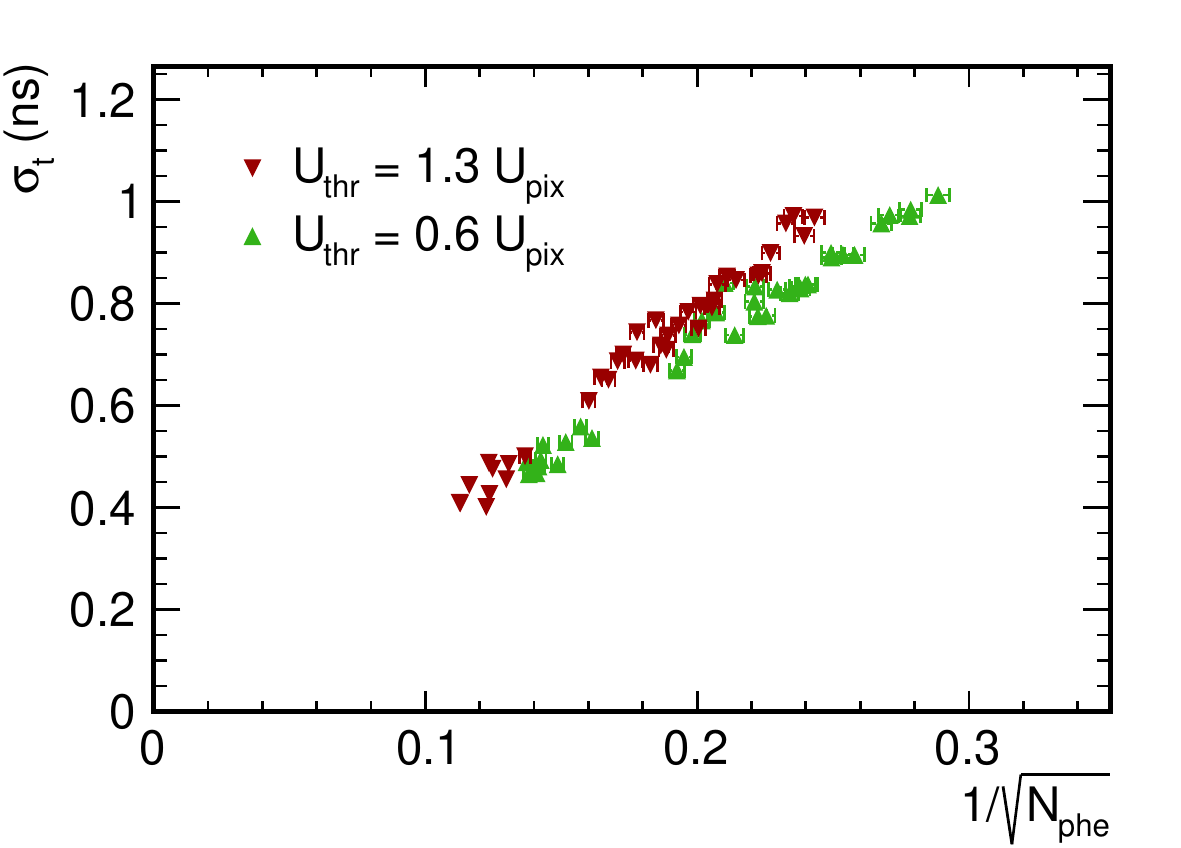}
   \caption{\label{fig:trend} Time resolution of a single end of a strip as a function of the inverse square root of the average number of photoelectrons for various muon positions and all tested strip configurations.}
\end{figure}

Results for the tested strips are summarized in Table \ref{tab:performance}. All resolution numbers show dependence on the light yield. \figref{fig:trend} shows the time resolution per strip end as a function of the inverse square root of the number of photoelectrons for various muon positions and all tested strip configurations. Data from measurements with different discriminator threshold levels $U_{\text{thr}}$ are shown in different colors. Observed linear dependence confirms the basic relationship determining the time resolution

\begin{equation}
\label{eq:sigmavsnpe}
  \sigma_t = \frac{ \sigma_0 }{ \sqrt{\ntot{} } }
\end{equation}

The parameter $\sigma_0$ is determined by the single photoelectron arrival time distribution and by the number of photoelectrons required to detect the timing signal ($U_{\text{thr}}$). The distribution of the arrival times of photoelectrons is determined by several parameters including the scintillation time, the fluorescence time of the WLS fiber, photon paths through the scintillator and the WLS fiber, intrinsic time resolution of the SiPM and the noise. Although strips $A$ and $B$ (design \subref{fig:minos}) use different scintillator material and geometry than the strip $C$ (design \subref{fig:bicron}), data from all strips for a given $U_{\text{thr}}$ appear on the same line, suggesting similar values of $\sigma_0$ for both strip designs. Superior resolutions achieved with design \subref{fig:bicron} are mainly due to the higher number of WLS fibers and thus more effective photon collection.

For $U_{\text{thr}} = 1.3 U_{\text{pix}}$, the value of $\sigma_0$ is somewhat higher, reflecting the fact that the distribution of the arrival times of the second photoelectron is somewhat wider than that of the first.

\begin{table}
\centering
\caption{\label{tab:sigma1pe} $\sigma_0$ for different strip designs and different $U_{\text{thr}}$.}
  \begin{tabular}{ l c c }
    \hline
    Design                              & $U_{\text{thr}} / U_{\text{pix}}$ & $\sigma_0$ (ns)  \\
    \hline
    \multirow{2}{*}{\subref{fig:minos}} &           0.6                     &   $3.57\pm0.02$ \\
                                        &           1.3                     &   $3.94\pm0.02$ \\
    \hline
    \multirow{2}{*}{\subref{fig:bicron}} &           0.6                    &   $3.44\pm0.04$ \\
                                        &           1.3                     &   $3.64\pm0.07$ \\
    \hline
  \end{tabular}
\end{table}

The average values of $\sigma_0$ calculated from the data points for the two strip designs and different $U_{\text{thr}}$ are listed in \tableref{tab:sigma1pe}. 
In comparison, the quadratic sum of the characteristic light-emission times of the scintillator and the WLS fiber is 3.3~ns \cite{BicronStrip, BicronWLS}.
The small difference includes effects such as the travel time distribution of photons in the scintillator and the WLS fiber, the time resolution of the SiPM and the electronic noise in the readout chain. We conclude that the measured time resolution is close to the ultimate resolution achievable with the used scintillators and WLS fibers and is mainly determined by the light-emission times of the scintillator and the WLS fiber. 

The range of angles for photon transport in the fiber necessarily creates a spread of photon travel times proportional to the distance covered by the photons. With further increase of strip length, it is expected that the spread in the photon transport times becomes comparable to $\sigma_0$ for short strips. 
A simplified Monte Carlo simulation of photon emission, reemission and transport times from the incident muon position was performed to estimate the length of a strip at which photon transport through the fibers begins dominating $\sigma_0$. The simulation assumes photons travelling along straight lines in the scintillator and the fiber materials with absorption probability obtained from the material datasheets. The light emission in the scintillator is assumed to be a two-stage cascaded deexcitation process. The scintillation time is thus obtained as a sum of samples from two exponential distributions with time constants equal to the \textit{rise time} and the \emph{decay time} of the scintillator material. The propagation of photons through the scintillator is simulated as a series of diffuse reflections from the reflective coating/wrapping with a small probability of capture at the WLS fibers at every passage across the location of the fibers. The time of photon reemission in the WLS fiber is sampled from the exponential distribution with time constant taken from the WLS fiber datasheet. 
Only photons satisfying the conditions for the total internal reflection in the WLS fiber are transported down the fiber.
The simulation confirms the near independence of $\sigma_0$ on the strip design and indicates that the independence of $\sigma_0$ on the strip length continues up to a strip length of $\sim 4\unit{m}$.

The time resolution and, consequently, the muon hit position resolution of the scintillator strips with WLS fibers and SiPM readout depends on a number of factors that can be split in two groups. The first group defines the distribution of arrival times of the photoelectrons and, thus, the $\sigma_0$. This group includes the following:
\begin{itemize}
  \item Scintillation time distribution
  \item The distribution of the photon travel paths through the scintillator, including reflections, before being captured on the WLS fibers
  \item The fluorescence time of the WLS fibers
  \item The time distribution of the photon propagation along the fiber 
  \item The time resolution of the SiPM
\end{itemize}
The second group of factors defines the photoelectron yield per muon, \ntot. This group includes:
\begin{itemize}
  \item The scintillator light output
  \item Geometry of the scintillator and the WLS
  \item Light capture and transport efficiency of the WLS. 
  \item Number of WLS fibers per strip
  \item Light attenuation
  \item Quality of the optical connections 
  \item Quantum efficiency of the SiPM
\end{itemize}

Improvement of the time resolution w.r.t.\ the results presented would require improvements of factors from either of these two groups. To improve the speed of the production and collection of the photons, timing characteristics of scintillators and WLS fibers would have to be improved. Possibilities in this area are limited. The scintillator and WLS fiber materials used in the present work are among the fastest available and performance has not changed much in the past two decades. Both studied designs use the maximum practical number of WLS fibers for light collection, limited either by the geometry of the strip, or by the active surface of the light detector. An improvement of time resolution by improving photoelectron yield is challenging due to the $1/\sqrt{\ntot{}}$ dependence of the resolution. We conclude that the resolutions measured in this study are close to the practical limits achievable with the present state-of-the-art materials and strip designs.

\section{Conclusions}
\label{sec:conclusions}

Prototype scintilator+WLS strip configurations with SiPM readout have been tested in the muon beam of the Fermilab Test Beam Facility. The strip with the maximum number of WLS fibers performs the best, yielding up to 137 photoelectrons per muon per strip. The time resolution of the best strip design is 330~ps and the position resolution is 5.4~cm. Results for strips up to 2~m in length show no deviation from the relationship $\sigma_t =  \sigma_0 / \sqrt{\ntot{}} $. 

Scintillator strips with WLS fibers and SiPM readout represent a robust and economic solution for large muon detection systems embedded in the return yoke of a collider detector, in which time resolution of a fraction of ns is required. The use of long strips allows the realization of large detectors with position resolution of a few cm with a moderate number of readout channels.

\section{Acknowledgments}
\label{sec:acknoledgments}

The authors acknowledge the support received from the Ministry of Education and Science and the National Research Center ``Kurchatov Institute'' (Russian Federation), from the Ministry of Education, Science and Technological Development  (Republic of Serbia) within the projects OI171012 and OI171018 and from the Department of Energy (United States of America).

\section*{References}

\bibliography{../bibliography/bibliography.bib}

\end{document}